%% file: main.tex
\documentclass[fontsize=13pt]{article}
\usepackage{graphicx}
\usepackage{amsmath}
\usepackage{amssymb}
\usepackage{txfonts}
\usepackage[]{hyperref}
\hypersetup{colorlinks, citecolor=blue, filecolor=black, linkcolor=black, urlcolor=blue,
            pageanchor=true}
\usepackage{siunitx}
\usepackage{textgreek}
\usepackage{natbib}
\usepackage[letterpaper, left=.8in, right=.8in, top=1in, bottom=1in]{geometry}
\usepackage{threeparttable} 
\usepackage{booktabs} 
\usepackage{multicol}

\newcommand{\des}{DES\-\mbox{TINY\hspace{-.2mm}$^{\scriptscriptstyle +}\mkern-2.5mu$}}
\newcommand{\amet}{\textalpha-meteoroid}
\newcommand{\bmet}{\textbeta-meteoroid}
\newcommand{\eg}{e.g.,}
\newcommand{\ie}{i.e.,}

\newcommand{\heos}{\mbox{HEOS-2}}
\newcommand{\gorid}{GO\-RID}
\newcommand{\au}{\astronomicalunit}
\newcommand{\kms}{\km\per\s}

\newcommand{\effSA}{\ensuremath{\Omega_{\mathrm{eff}}}}
\newcommand{\earthRad}{\ensuremath{\textit{R}_\oplus}}
\newcommand{\geomFac}{\ensuremath{G}}
\newcommand{\nomArea}{\ensuremath{A_0}}

\newcommand{\kp}{\textit{Kp}}

\DeclareSIUnit\day{\text{day}}
\newcommand{\FetalAbbr}{{\hyperlink{Fechtig1979}{\textcolor{blue}{F79}}}}

\usepackage{etoolbox} 
\AtBeginEnvironment{thebibliography}{\small}

\begin{document}

\input{0_abstract}
  \input{1_introduction}

  \input{2_review}

  \input{3_discussion}

  \newpage
	\bibliographystyle{elsarticle-harv}
	\bibliography{references} 
  
\end{document}

%% file: 0_abstract.tex
\title{The unresolved mystery of dust particle swarms\\within the magnetosphere}

\author{Max Sommer$^{1,2}$}

\date{  \small{$^{1}$University of Stuttgart, Germany\\
        $^{2}$University of Cambridge, UK}\\[2ex]
        13 May 2024}
\maketitle

\begin{abstract}
  Early-generation in-situ dust detectors in near-Earth space have reported
  the occurrence of clusters of sub-micron dust particles that seemed unrelated
  to human spaceflight activities. 
  In particular, data from the impact ionization detector onboard the \heos{} satellite indicate 
  that such swarms of particles occur throughout the Earth's magnetosphere up to altitudes
  of \qty{60000}{\km}---far beyond regions typically used by spacecraft.
  Further account of high-altitude clusters has since been given by the GEO-deployed 
  \gorid{} detector, however, explanations for the latter have so far only been sought in 
  GEO spaceflight activity.
  
  This perspective piece reviews dust cluster detections in near-Earth space,
  emphasizing the natural swarm creation mechanism conjectured 
  to explain the \heos{} data---that is, the electrostatic disruption of meteoroids.
  Highlighting this mechanism offers a novel viewpoint
  on more recent near-Earth dust measurements.
  We further show that the impact clusters observed by both \heos{} and \gorid{}
  are correlated with increased geomagnetic activity.
  This consistent correlation supports the notion that both sets of observations
  stem from the same underlying phenomenon
  and aligns with the hypothesis of the electrostatic breakup origin.
  We conclude that the nature of these peculiar swarms remains highly uncertain,
  advocating for their concerted investigation by forthcoming dust science endeavours,
  such as the JAXA/DLR \des{} mission.
\end{abstract}

%% file: 1_introduction.tex
\section{Introduction}

Since the first reliable and highly sensitive in-situ dust detectors were deployed in the 1970s,
the clustering of dust particles, that is, the occurrence of multiple impacts on timescales of minutes or hours,
has been a commonly observed phenomenon in near-Earth space.
In particular, it was the impact ionization detectors 
onboard the satellites Prospero and \heos{} that gave the first conclusive account of such clusters 
\citep{Bedford1973earth,Hoffmann1975first}.
This detector type registers the plasma generated in hypervelocity impacts of micron and sub-micron particles,
and posed a major improvement over previous designs, such as the microphone-type sensors,
which proved to be highly susceptible to the detection of spurious events \citep{McDonnell1978microparticle}.
The detected clusters were interpreted as spatially confined swarms of micrometeoroids, 
encountered by the spacecraft.
Such formations of particles would be expected to rapidly disperse,
and thus, an active mechanism that continuously created these clusters would be required.
This led \citet{Fechtig1976nearearth} to propose that the clusters are generated
by the fragmentation of larger meteoroids, for instance, 
through phase-change-induced or electrostatically-driven disruption
triggered upon entering the ionosphere.
Analysing the \heos{} dust counter data, which covered a broad range of 
distances from Earth due to the satellites highly elliptic orbit, 
\citet{Fechtig1979micrometeoroids} (in the following referred to as F79) 
conclude that rather confined particle swarms, 
presumably generated by the electrostatic breakup of fluffy meteoroids, 
occur throughout the Earth's magnetosphere, that is,
up to altitudes of roughly \qty{60000}{\km}.
Further account of clustering was given by several other dust counter instruments, 
in particular in low Earth orbit (LEO), such as those onboard the Long Duration Exposure Facility (LDEF, launched 1984)
and other satellites (see Section~\ref{SECT:LEO}),
as well as at altitudes beyond LEO by the Munich Dust Counter (MDC, 1990)
and the Geostationary Orbit Impact Detector (\gorid, 1996).

Some clusters detections during these later missions could be attributed to human spaceflight activities,
in particular, to specific firings of solid rocket motors (SRMs) \citep[\eg][]{Schobert1997al2o3}.
SRMs produce large amounts of solid combustion products
in the form of micron-sized aluminium oxide particles,
which, if used for geostationary orbit (GEO) insertions or retrograde LEO deorbiting manoeuvres,
can create circumterrestrial streams of micro-debris with lifetimes of up to months 
\citep{Bunte2003detectability,Stabroth2008influence}.
As the use of orbital SRMs ramped up in the late 1970s and 1980s 
\citep[to enable utilization of the GEO, see \eg][]{McDowel1997kick,Wegener2004population},
they gained attention for their potential to cause hazard to spacecraft 
through their generated micro-debris trails \citep{Mueller1985effects,Akiba1990alumina}---amidst
the growing awareness of the space debris problem in general
\citep{Kessler1984orbital,Kessler1991collisional}.
In fact, the analysis of impacts on retrieved spacecraft surfaces showed that, in LEO, the
debris dust flux dominates over the natural dust flux 
\citep{Laurance1986flux,Graham2001microparticle,Horz2002metallic}. 

Following this increased scrutiny of dust sources from human activity,
clustered impacts recorded by the GEO-deployed \gorid{} were generally attributed to micro-debris
\citep{Drolshagen2001measurementsb,Drolshagen2001measurementsa},
even though only few such events could be directly linked to SRM firings.
It appears that earlier hypotheses regarding natural particle swarms occurring
throughout the magnetosphere, as indicated by \heos{} data---remarkably, recorded prior to the extensive
utilization of the GEO---were not considered in the analysis of the \gorid{} data.
However, aspects of \heos{} and \gorid-detected clusters suggest that both studies
may have observed the same phenomenon.
In this perspective, we first review the measurements of particle clusters
in near-Earth space in general in Section~\ref{SECT:measurements}.
In Section~\ref{SECT:discussion}, we then reassess the observations of \heos{} and \gorid{} 
of clusters observed beyond LEO (\ie{} those by \heos{} and \gorid{}),
under the premise that both instruments have observed the `magnetospheric swarms'
postulated by \FetalAbbr.
We also provide an outlook for the future investigation of this phenomenon
by the upcoming \des{} mission.

%% file: 2_review.tex
\section{Measurements and interpretations of near-Earth dust clusters} \label{SECT:measurements}

\subsection{At LEO altitudes} \label{SECT:LEO}
The multitude of dust-counter-type instruments deployed in LEO gathered extensive evidence
for clustered impacts of submicron and micron-sized particles at such altitudes.
The first such account was given by the impact ionization detector onboard the Prospero satellite,
which registered large variations in the daily impact rate,
suggesting that around 64\% of impacts occurred in clusters 
\citep{Bedford1975flux,Bedford1975observations}.
In search for explanations, there was cautious speculation that meteoroids grazing 
the Earth's atmosphere could break up and shed a swarm of sub-micron fragments,
which would then be intercepted by the spacecraft \citep{Bedford1975flux}.
(This mechanism was later ruled out for swarms observed at higher altitudes, see Section~\ref{heos2}.)
Similarly, the capacitor-type dust detector onboard Explorer 46 reported variations
in the daily impact rate, which were attributed to the presence of submicron particles 
within known meteor showers \citep{Singer1980submicron}---contrary
to the theoretical expectation of micro-particle-depleted streams
due to the action of radiation pressure \citep[\eg{}][]{Dohnanyi1972interplanetary}.

Conclusive evidence for the occurrence for compact particle clusters in LEO
was eventually given by the high-time-resolution dust counters onboard LDEF,
which occasionally recorded bursts of impacts with instantaneous rates 
up to four orders of magnitude higher than the mean rate \citep{Oliver1995ldef}.
Some of these bursts were seen repeatedly at certain points along the spacecraft orbit,
indicating intersections with distinct circumterrestrial dust streams, thus interpreted
to stem from human activity \citep{Cooke1995orbital}.
Two of those repeating clustering events could then indeed be linked to the exhaust dust streams
of specific SRM firings \citep{Schobert1997al2o3,Stabroth2007explanation}.
However, the origin of most of the detected clusters remained unresolved.

Further confirmation that the majority of dust impacts in LEO occurs in clusters, 
some of which representing circumterrestrial streams, has since been given by
the dust-counter-type instruments SPADUS \citep{Tuzzolino2005final}, 
DEBIE-1 \&~2 \citep{Schwanethal2005analysis,Menicucci2013inflight}, 
and SODAD-1 \&~2 \citep{Durin2009sodad,Durin2016system,Durin2022active}.
One of the streams detected by SPADUS could be attributed to a known SRM firing
\citep{Neish2004numerical,Bunte2005detection}, whereas another one 
appeared to be associated with the explosion of a launch vehicle upper stage
\citep{Tuzzolino2001insitu}.

\subsection{Beyond LEO}
While the occurrence of clusters in the densely-populated LEO region may plausibly
be the result of human activity in a number of ways
(\ie{} collisions and explosions of satellites or debris, orbital SRM firings, 
substances released by crewed spacecraft, etc.),
the occurrence of clusters at higher altitudes is more difficult to explain.
Nonetheless, evidence for their existence has been gathered by a few in-situ dust detectors,
which are reviewed in the following.

\subsubsection{HEOS-2} \label{heos2}
\begin{figure}[h!tb]
  \centering
  \includegraphics[width=.5\linewidth]{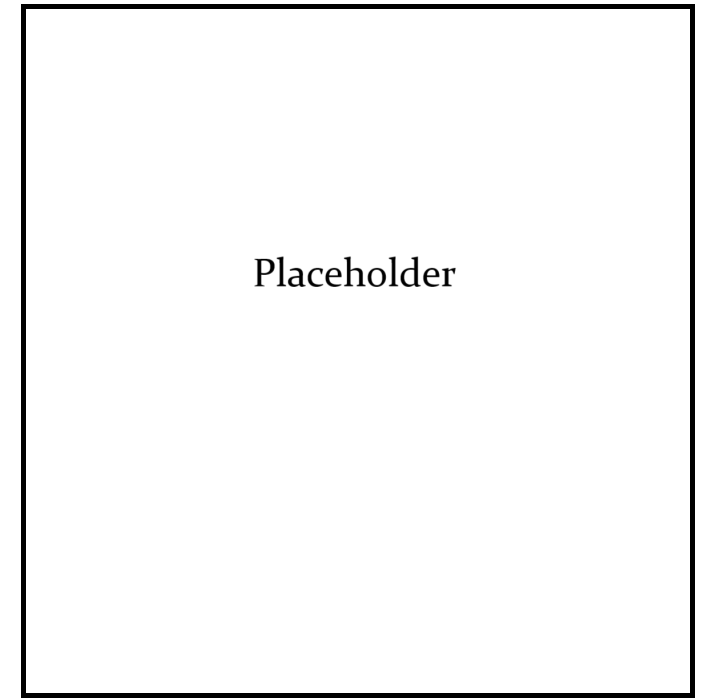}
  \caption{
    Illustration of the \heos{} orbit through the plasma and field regions of the Earth's magnetosphere.
    Reproduced from \citet{Fechtig1979micrometeoroids}. (Due to copyright restrictions this figure
    is only included in the journal version of this article.)\protect\footnotemark[1]} 
  \label{fig:HEOS_orbit}
\end{figure}

The first and most extensive account of particle clusters at altitudes beyond LEO was given
by the dust counter onboard \heos{}.
The \heos{} spacecraft, launched in 1972, carried an impact ionization detector 
on a highly eccentric orbit with varying perigee in a range of 
\qtyrange[range-units=single,range-phrase=--]{350}{3000}{\km} 
and apogee of \qty{240000}{km}, as illustrated in Figure~\ref{fig:HEOS_orbit}.
Due to that orbit, \heos{} could sample the dust flux over a wide range of altitudes, 
from an almost `pristine' interplanetary background flux
around its apogee, all the way to the near-Earth dust flux at only a few \qty{1000}{\km} altitude.
Besides an interplanetary sporadic flux described by random impacts,
the \heos{} data indicated the presence of two types of cluster phenomena:
(1) the `swarms', which were characterized by short intervals between consecutive impacts of up to 15~minutes, and
(2) the `groups', with longer intervals of up to 12~hours \citep{Hoffmann1975first,Hoffmann1975temporal,Fechtig1979micrometeoroids}.

\footnotetext[1]{Journal version of this article: \href{{https://doi.org/10.1098/rsta.2023.0370}}{{https://doi.org/10.1098/rsta.2023.0370}}}
\setcounter{footnote}{1}

The swarms were exclusively encountered at distances of up to 10~Earth radii (\si{\earthRad}), 
which is about the extent of the Earth's magnetosphere dipole field (about twice the GEO altitude).
These `magnetospheric swarms' generated an averaged particle flux rate
about one order of magnitude above the sporadic background.
Examining different sensor viewing directions perpendicular to the Sun---namely,
the Earth apex, antapex, ecliptic north, and south---\heos{}
also revealed that the swarms occurred anisotropically,
with a slight preference for the Earth apex direction (\FetalAbbr).
The derived masses of the individual particles ranged mostly within \qtyrange{e-14}{e-12}{\gram},\footnote{
  Corresponding to particle radii of about \qtyrange[range-units=single,range-phrase=--]{100}{500}{\nm}, 
  or $\beta$ factors of in the order of 0.5--1.
} similar to the largest of the hyperbolic \bmet{}s 
irradiating from the Sun \citep[\eg][]{Wehry1999identification},
which however the \heos{} sensor was insensitive to, due to its pointing.
Impact speeds of the swarm particles derived by rise time analysis of the impact charge signal
were found to be in the order of \qtyrange[range-units=single,range-phrase=--]{4}{16}{\kms}. 

\FetalAbbr{} show that the swarms could be caused by the breakup of fluffy interplanetary
meteoroids that fragment due to inner, repulsive electrostatic forces 
\citep[a mechanism first described by][]{Opik1956interplanetary}, resulting from being 
charged up to high potentials while traversing the Earth's magnetosphere.
From the number of registered impacts and the total encounter duration of each detected swarm, 
they estimate the total numbers of particles per swarm to fall in the range from \numrange{e14}{e19}, 
corresponding to individual parent meteoroid masses between \qty{10}{\gram} and \qty{1000}{\kilo\gram} 
(\FetalAbbr), with a geometric mean of \qty{5.2}{\kilo\gram}.
(Note that the largest observed swarm, corresponding to a progenitor of nearly 
\qty{1000}{\kilo\gram}, appears to be divided into four sub-swarms, 
suggesting the breakup of multiple smaller objects in close temporal proximity.)
Such meteoroids should be observable as visible bright meteors, also called `fireballs', when entering the atmosphere.
The meteoroids producing the Ceplecha type~III fireballs, which show high ablation ability
and are thus thought to stem from low-bulk-density meteoroids \citep{Ceplecha1977meteoroid,Ceplecha1998meteor},
are considered by \FetalAbbr{} the most likely candidates for the swarm progenitors.
These low-density-type meteors, however, are associated with cometary source orbits
that generally produce high atmospheric entry velocities 
(\qtyrange[range-phrase=--,range-units=single]{20}{70}{\kms}),
which is incompatible with the rather low-velocity swarms.
\FetalAbbr{} acknowledge a lack of low-velocity type~III fireballs,
which could represent the sought-after swarm progenitors, 
but note that this apparent deficiency could potentially be explained
by selection effects in meteor observations as well as swarm production.
Based on the swarms' directionality, they rule out creation via aero-fragmentation 
(the speculated origin of the Prospero clusters, see Section~\ref{SECT:LEO}) 
or a relation to meteor showers \citep{Dohnanyi1977micrometeoroid,Fechtig1979micrometeoroids}.

The more spread-out `groups', on the other hand, were detected at all distances from Earth
(\ie{} up to the satellite's apogee of around \qty{240000}{\km}),
yet crucially, they appeared almost exclusively when the Moon was within the sensor's field of view
\citep{Hoffmann1975first,Fechtig1979micrometeoroids}. 
Backtracing of the group particles' trajectories further substantiated the idea
of a lunar origin \citep{Hoffmann1975temporal}.
Consequently, the authors interpreted these groups as originating from 
occasional larger meteoroid impacts on the Moon
that would create significant amounts of ejecta escaping lunar gravity, subsequently 
roaming through the Earth-Moon system in somewhat dispersed formations of particles.
By modelling the motion of escaping ejecta clouds through the Earth-Moon system,
it was further concluded that their expected spatial dimensions are consistent
with \heos{}'s encounter times of the groups, and that lunar impactors in the order
of \qty{1}{\kg} could suffice to create the observed group number densities \citep{Dohnanyi1977groups}.
Attempts to correlate the occurrence of groups with lunar impact events sensed by 
Apollo-deployed seismometers were inconclusive, due to the much higher number 
of seismic events compared to group detections (\FetalAbbr).
In this paper, we will not consider these lunar-origin clusters further, 
and instead focus on the near-Earth swarms observed by \heos{}.

\subsection{Munich Dust Counter}
The findings of \heos{} are reinforced by the data of the impact-ionization-type 
Munich Dust Counter (MDC) onboard Hiten, launched in 1990.
As \heos{}, the Hiten spacecraft was placed on a highly eccentric orbit
with initial perigee of a few \qty{1000}{\km} and apogee of \qty{300000}{\km}.
MDC, similarly to the \heos{} detector, reported a ten-fold increase of flux in regions 
near Earth (closer than \qty{100000}{\km}) compared to interplanetary space and
an anisotropy in the flow direction, with a preference for the Earth apex direction
\citep{Iglseder1993analysis,Iglseder1993cosmic}.
In addition, strong clustering with instantaneous particle rates of up to five orders of magnitudes 
over the mean flux was observed.
Unfortunately, however, no further details about the MDC-detected clusters have been published.

\subsection{GORID}
The Geostationary Orbit Impact Detector (GORID) was a refurbished engineering model unit
essentially identical to the impact ionization dust detectors flown 
on the interplanetary probes Ulysses and Galileo \citep{Grun1992galileo,Grun1992ulysses}.
GORID was mounted on the Russian telecom satellite Express-2, which was placed in GEO in 1996
\citep{Drolshagen1999microparticles}.
\gorid{} data indicated a significant excess of clustered impacts, which were characterized
by separations in time from minutes to an hour, 
dominating over random impacts by a factor of about four
\citep{Drolshagen2001measurementsb,Graps2007geo}.
Similarly to the case of LDEF, few clusters were observed recurrently at certain points along the orbit,
and could be dynamically linked to the exhaust dust streams of specific SRM firings 
\citep{Bunte2003detectability,Bunte2005detection}.
The vast majority of detected clusters, however, were non-recurrent 
and could not be attributed to any specific source.
This led \citet{Bunte2005detection} to conjecture 
an unknown micro-debris-cloud-generating mechanism present at GEO altitudes.
In that sense, it was speculated that the breakup of larger SRM slag particles
via electrostatic fragmentation could act as such a mechanism 
\citep{Graps2007geo}.\footnote{
  SRM slag particles are larger (up to cm-sized) aluminium-oxide clumps ejected at the 
  end of an SRM burn with low relative velocity \citep{Jackson1997historical}.
  Fragments of slag produced in GEO insertions (consequentially exhibiting near-GEO orbital parameters)
  would have too-low velocities relative to \gorid{} (few \qty{100}{\m\per\s}) to be efficiently detected
  via impact ionization.
  However, GTO insertions would produce slag with velocities relative to \gorid{}
  of \qtyrange[range-units=single,range-phrase=--]{1.5}{2}{\kms}.
}

Ultimately, all \gorid{}-detected clusters were attributed to debris-related phenomena 
\citep{Drolshagen2001measurementsb,Graps2007geo}
with no consideration of the possibility that the clusters are of natural origin.
One reason for this is that the impact velocities inferred from charge signal rise times
indicate most impactors to have slow speeds (\qty{<5}{\kms}), which would be compatible with 
relative speeds of debris in GTO-like orbits.
However, it is noted that these impact speeds had been found to be unreliable
for a number of reasons.
In particular, impacts on the insides of the detector's side walls instead on the impact target
would generate events with abnormally long rise times, and thus appear as slow, even though they are not
\citep[see also][]{Stubig2002new,Altobelli2004influence,Willis2004influence,Willis2005decreased}.
Later analysis of the \gorid{} data therefore ignores the rise-time-derived speeds 
altogether \citep{Graps2007geo}.
By geometrical consideration, one might expect that around half of the impacts
inside the detector occur on the side walls (assuming an isotropic flow).\footnote{
  The effective solid angle \effSA{} of \gorid{} including its inner side walls
  can be approximated by an unobstructed plate detector with the same nominal sensitive area
  (all particles entering the instrument are detected), 
  which has $\effSA\,=\,\pi\,\mathrm{sr}$.
  The $\effSA$ of the \gorid{} impact target is \qty{1.45}{sr} \citep{Grun1992galileo}.
  Thus, the $\effSA$ of just the inner walls (\qty{1.69}{sr}, calculated as the remainder of the two)
  is comparable to that of the impact target.
}

One striking feature of the \gorid{}-observed clusters is the strong modulation of their occurrence
with the local time of the satellite.
Clusters were detected with a strong preference around local midnight,
that is, when the satellite was on the Sun-opposing side of the Earth, 
and the spacecraft-body-fixed sensor was sensitive to a flux from the Earth apex direction
\citep{Bunte2005detection,Graps2007geo}.
For the half of the orbit, when the sensor was pointing away from apex, virtually no clusters were detected.
There is no obvious reason why debris would preferentially be located around local midnight,
although \citet{Graps2007geo} note that an unknown process occurring only in the magnetotail region might
facilitate debris cluster detections.

Another noteworthy feature is that the distribution of the impact-generated
ion grid charges (`$Q_\mathrm{i}$') shows a greater tendency towards smaller charges
for clustered impacts compared to random events \citep{Graps2007geo}.
This trend could imply a steeper mass distribution, 
possibly indicating an increased prevalence of smaller particles within clusters.

Lastly, measurements of \gorid{}'s charge-senstive entry grids indicated
that cluster particles were highly negatively charged \citep{Bunte2006processing,Graps2007geo},
as opposed to the positive equilibrium potential particles are thought to assume
in interplanetary space as well as inside the magnetosphere under average conditions \citep{Horanyi1996charged}.
Positive charges, however, occurred only among the randomly impacting particles, 
according to the \gorid{} measurements.

%% file: 3_discussion.tex
\section{Discussion} \label{SECT:discussion}

\subsection{\heos{} and \gorid{} clusters - a common origin?} \label{SECT:Swarms}
Having reiterated the main findings about dust particle clusters in near-Earth space, 
we will now reassess them in context to each other.
In particular, we will attempt to reconcile the high-altitude clusters observed by \heos{} and \gorid{},
which have garnered notably different interpretations.
The fact that both instruments reported an apparent anisotropy favouring the Earth apex,
and typical cluster encounter durations of up to \qty{90}{minutes} suggests
that they are related to each other, and thus, that they could stem from the same phenomenon.
In the following, we will discuss several aspects that could support this notion---specifically,
\FetalAbbr{}'s hypothesis of the magnetospheric swarms---and
weigh in on potential swarm progenitor candidates,
as well as other explanations for the observed clusters.

\subsubsection{Anisotropy} \label{SECT:Anisotropy} 
\begin{figure}[h!tb]
  \centering
  \includegraphics[width=.7\linewidth,trim={0 0 0 5mm},clip]{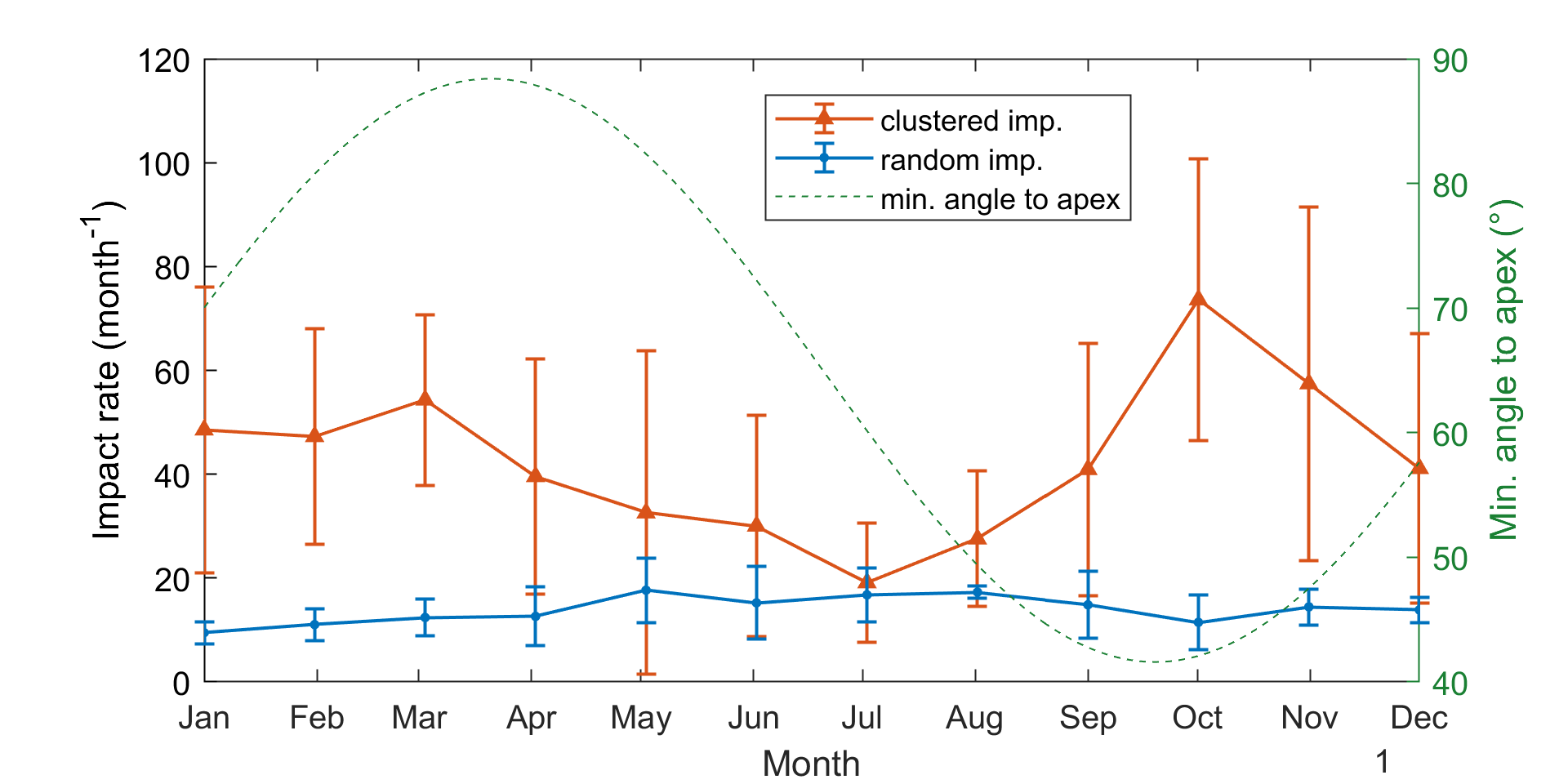}
  \caption{
    Monthly mean impact rates recorded by \gorid{}
    per calendar month averaged over the entire observing period (1997--2002).
    Error bars represent the year-to-year standard deviation.
    Based on data from \citet{Drolshagen2006insitu}.
    Also shown is the minimum angle between the sensor boresight and the Earth apex direction
    (reached once per day at local midnight) over the course of the year.
  }
  \label{fig:GORID_monthly_year}
\end{figure}

Both instruments reported preferential detection of clusters when exposed to the Earth apex direction,
although the observed anisotropy was arguably more pronounced for \gorid{} than for \heos{}
(compare, \eg{} \FetalAbbr{}~Tab.~7 with \citet{Graps2007geo}~Fig.~2).
Interestingly, although the \gorid{} data indicate a strong preference for cluster detection when exposed
to the Earth apex, the effective sensitivity towards that direction remained relatively low.
Due to \gorid{}'s body-fixed mounting on the GEO satellite with a boresight pointing only \qty{25}{\degree}
away from equatorial north, the minimum angle between boresight and the apex direction
(which was achieved once per day, during local midnight on the geostationary orbit)
varied between \qtyrange[range-phrase={ and }]{42}{88}{\degree}, 
depending on the time of the year.
Thus, considering \gorid{}'s angular sensitivity profile \citep{Grun1992galileo}, 
one would expect a distinct seasonal modulation of cluster detections,
if the clusters were indeed coming preferentially from the apex direction.
The maximum cluster occurrence rate should occur in autumn when the sensor's exposure
to the Earth apex direction is highest (during local midnight hours),
whereas the minimum rate should occur in spring. 
To check this proposition, the mean flux of clustered events during each calendar month
averaged over the mission duration is shown in Figure~\ref{fig:GORID_monthly_year},
alongside the minimum angle between sensor boresight and the apex direction.
Although the highest rates are indeed observed in autumn, 
the minimum detection rate of clusters occurs in summer.
Thus, a clear correlation of the cluster occurrence rate with the 
sensor's sensitivity toward the Earth apex direction is not evident.

A reduced modulation of rates with the exposure to the apex
could be due to the sensor's asserted sensitivity to wall impacts 
\citep{Stubig2002new,Altobelli2004influence}, 
which considerably widens the angular sensitivity profile.
On the other hand, the lack of a clear modulation throughout the year 
in combination with the observed strong modulation with local hour,
indicates that it is not the sensor's exposure to the apex direction 
that is the crucial factor for cluster detection,
but rather the location of the spacecraft on the tail side of the magnetosphere.
The tail side could indeed be preferential for the detection of the putative 
magnetospheric swarms, due to the proximity to the magnetosphere's plasmasheet.
The plasmasheet, with its elevated electron fluxes,
had been suggested by \FetalAbbr{} as another potential swarm genesis region,
besides the field line regime connected to the auroral zones traversed by \heos{} (\FetalAbbr{}).

\subsubsection{Geomagnetic activity} \label{SECT:geom_activity} 
Considering that \FetalAbbr{} identified high fluxes of energetic electrons (\qty{>10}{keV})
in the surrounding plasma as a crucial factor for charging up meteoroids enough
to cause their electrostatic breakup (\FetalAbbr{}),
it is reasonable to assume that the occurrence of swarms is correlated with geomagnetic activity.
Plasma motion and electric currents inside the magnetosphere are enhanced during geomagnetic storms,
which are driven by solar wind disturbances, which in turn are produced by solar activity 
\citep{Richardson2000sources}.
We may thus analyse the occurrence of clusters registered by \heos{} and \gorid{}
in the context of the solar cycle timing and the geomagnetic activity level.

\begin{figure}[!htb]
  \centering
  \begin{tabular}{cc}
  \raisebox{27mm}{\heos:}  & \hspace{-5mm} \includegraphics[width=.8\linewidth,trim={0 0 0 5mm},clip]{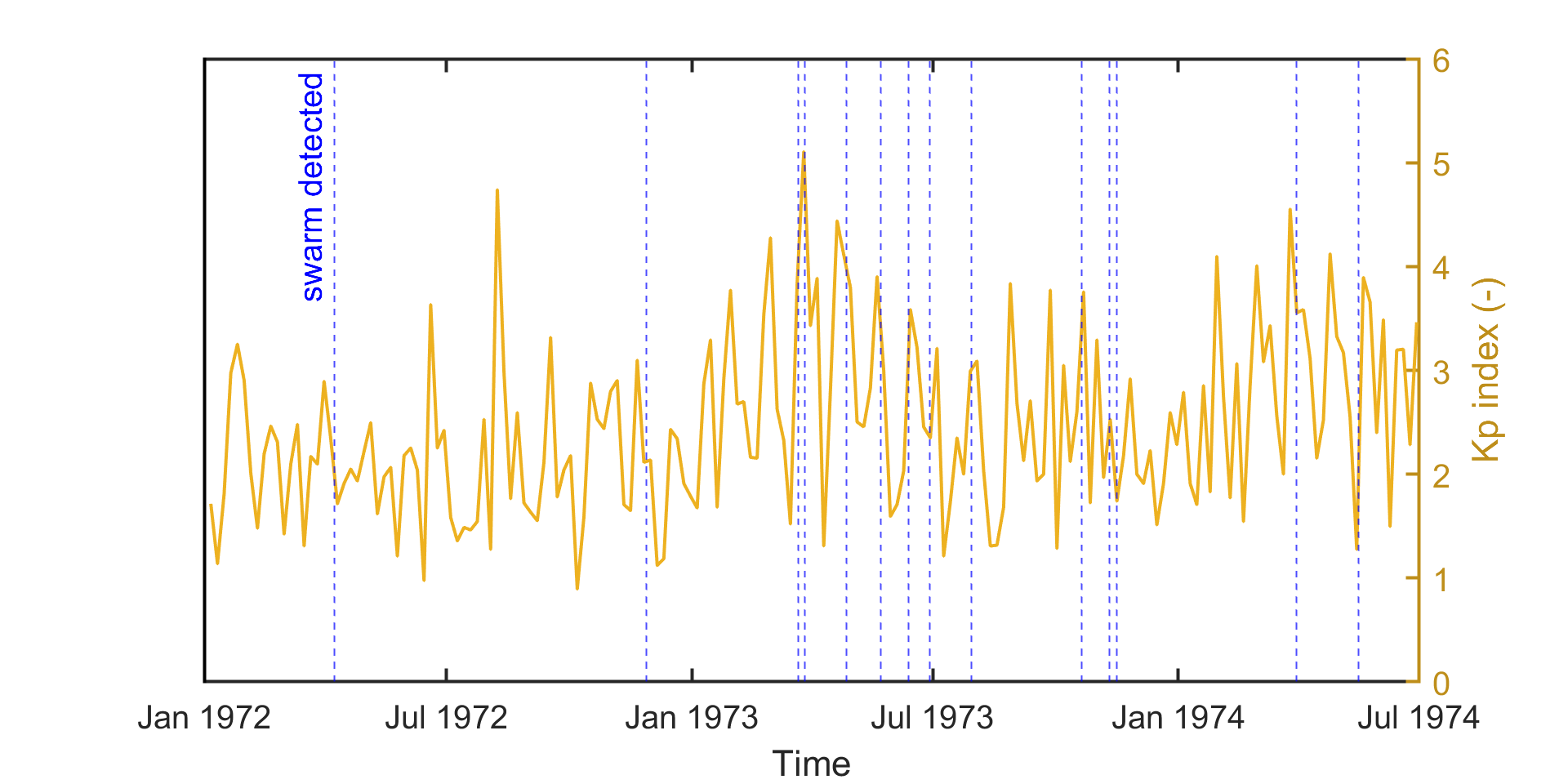}\\
  \raisebox{27mm}{\gorid:} & \hspace{-5mm} \includegraphics[width=.8\linewidth,trim={0 0 0 5mm},clip]{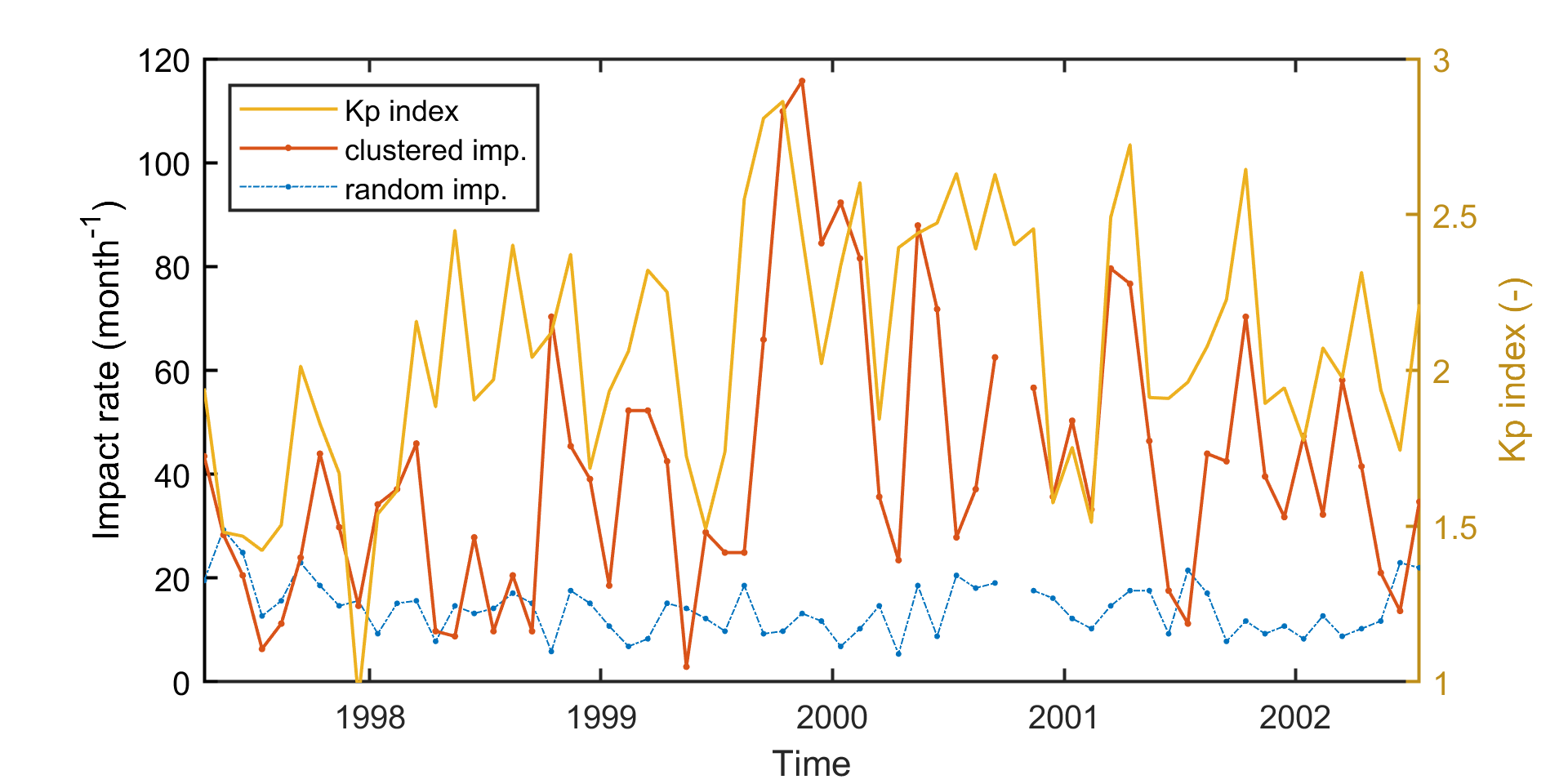}
  \end{tabular}
  \caption{
    Cluster occurrence with respect to geomagnetic activity.
    (Top) Times of swarm detections by \heos{}
    over the entire observing period (obtained from F79),
    indicated by vertical blue lines.
    Also shown is the average \kp{} index for 5-day intervals,
    indicating the level of geomagnetic activity.
    (Bottom) Monthly mean detection rates recorded by \gorid{} of random and clustered impacts
    over the entire observing period (obtained from \citet{Drolshagen2006insitu},
    considering only `class~3' events, supposed to be real impacts). 
    Also shown is the average \kp{} index for each month.
    \kp{} index data for both plots was obtained from \citet{Matzka2021GFZ,Matzka2021geomagnetic}.
  }
  \label{fig:HEOS_Kp_mission}
\end{figure}

The observation period of \heos{} (1972--1974) occurred well after the peak of solar cycle 20 (1969),
and during the cycle's decline (minimum in 1976).
A cycle's declining phase is typically not associated with major geomagnetic storms,
due to the less frequent occurrence of coronal mass ejections (CMEs).
The decline of solar cycle 20, however, was marked by a period of exceptionally 
high geomagnetic activity occurring in 1973--1975, 
caused by persistent high-speed solar wind streams
\citep{Gosling1977unusual,Richardson2000sources,Richardson2012solar}.
High fluxes of energetic electrons, in particular, are associated with high-speed streams 
\citep{Reeves2011relationship,Boynton2013analysis}.
Notably, \heos{} detected 13 of the 15 swarms in 1973 and 1974,
even though this period corresponded to only 64\% of the observation time.
To test whether the occurrence of particle swarm events correlates with elevated geomagnetic activity,
we can analyse the `\kp{} index' preceding these events. 
The \kp{} index, derived from geomagnetic observatory data at a 3-hour interval, 
provides a standardized measure of global geomagnetic activity levels on a scale from 0 to 9, 
with higher values indicating increased disturbance, 
and has been recorded for more than 70~years \citep{Matzka2021geomagnetic}.
Our analysis shows that, compared to the overall mean \kp{} index
of 2.45 (SD = 1.38) over the mission duration,
the mean of the \kp{} indices directly preceding a \heos{} swarm detection
is notably higher at 3.46 (SD = 1.19).
This difference is statistically significant, as supported by the obtained 
\textit{p}-value of 0.005 from a two-sample \textit{t}-test, suggesting that there is only
a 0.5\% probability of observing such a difference in means if there was no correlation.
These findings indicate a preferential association between particle swarm detections 
and heightened geomagnetic activity, as reflected by the \kp{} index.
This is also illustrated in Figure~\ref{fig:HEOS_Kp_mission} (top), 
which shows the timings of \heos{} swarm detections
next to the \kp{} index (averaged for 5-day intervals).

The observation period of \gorid{} (1997--2002) covers the rising phase and maximum 
of solar cycle 23 (min.\@ in 1996 and max.\@ in 2001).
In the case of \gorid{}, our dataset is limited to the monthly mean detection rate of clustered impacts
\citep{Drolshagen2006insitu}, rather than the precise timings of cluster detections.
Therefore, we compare the cluster impact rate also to the monthly mean \kp{} index, 
as depicted in Figure~\ref{fig:HEOS_Kp_mission} (bottom).
This comparison reveals a correlation between the cluster detection rate 
and the geomagnetic activity level, quantified by the Pearson correlation coefficient $r$.
While for the random event rate we obtain $r\,=\,-0.11$ (indicating no correlation), 
the cluster rate shows a moderate correlation with $r\,=\,0.57$.

It is important to note that the electron fluxes might not always be directly correlated with the \kp{} index.
In particular, the pumping of fluxes of relativistic electrons (\qty{>1}{MeV}) by geomagnetic storms
is known to be time-delayed and can cause elevated levels lasting for several days after a storm
has subsided \citep{Baker1998strong,Friedel2002relativistic,Boynton2013analysis}.
Conditions facilitating swarm formation may thus persist beyond geomagnetic activity peaks.
Nevertheless, the fact that both \heos{} and \gorid{} detected clusters preferentially 
during periods of high geomagnetic activity further hints at a common phenomenology.

The notion that swarm creation is associated with high geomagnetic activity is also consistent 
with the findings of \citet{Graps2000properties}, who modelled the equilibrium potentials
of grains within the magnetosphere for `quiet' and `active' conditions, 
as well as for different particle materials.
They conclude that, while quite conditions usually generate positive potentials (\qty{<15}{\volt}),
disturbed conditions can produce highly negative potentials (\qty{<-1000}{\volt})
for materials with low photoelectron yield.

\subsubsection{Impact rates} \label{SECT:Impact_rates} 
We may also compare the incidence rates of clustered impacts between \heos{} and \gorid{}.
Table~\ref{tab:SensGeom_compareHeosGorid} summarizes the key geometrical characteristics of both sensors,
the nominal sensitive area \nomArea{}, the maximum entry angle $\theta_{\mathrm{max}}$, 
the effective solid angle \effSA{}, and the geometric factor \geomFac{},
which is the product of \nomArea{} and \effSA{}, and represents 
the effective gathering power of an instrument with respect to an isotropic flux.
These parameters allow us to relate the impact rates onto \heos{} and \gorid{} to one another.
For a collimated dust stream and a directly upstream-pointing sensor, 
only the nominal sensitive area \nomArea{} is relevant,
such that \gorid{} would be exposed to about 10 times as many impacts as the \heos{} detector.
For an isotropic flow of dust (or a randomly varying instrument pointing) 
the geometric factor \geomFac{} is the relevant characteristic,
such that \gorid{} would be exposed to about 15 times as many impacts as \heos{}.

\begin{table}[!htb]
  \centering
  \caption{Comparison of sensor geometries.}
  \label{tab:SensGeom_compareHeosGorid}
  \begin{threeparttable}[b]
    \begin{tabular}{l S S[table-format = 3.1] S[table-format = 1.3] S[table-format = 4.0]}
      \toprule
      Instrument & \nomArea        & $\theta_{\mathrm{max}}$ & \effSA          & \geomFac                  \\
                 & \si{\square\cm} & \si{\degree}            & \si{\steradian} & \si{\square\cm\steradian} \\
      \midrule
      \gorid{} \citep{Grun1992galileo}     & 1000     & 70     & 1.45         & 1450 \\
      \gorid{} incl.\@ walls (approx.)     & 1000     & 90     & 3.14         & 3140 \\
      DDA \citep{Sommer2024science}        & 300      & 45     & 0.48         & 144  \\
      \heos{} \citep{Hoffmann1975temporal} & 95.4     & 60     & 1.03         & 98   \\
      \bottomrule
    \end{tabular}
  \end{threeparttable}
\end{table}

With \heos{}, a total of 207 swarm particles (split among 15 swarms) have been identified 
during an accumulated observation time below \qty{10}{\earthRad} of \qty{70}{days} 
(\FetalAbbr{}).
This amounts to an average incidence rate of \qty{2.9}{\per\day}.
On the other hand, \gorid{} detected a total of 2477 clustered events during a total observation time
of 1827~days \citep{Drolshagen2006insitu,Graps2007geo}, amounting to a mean rate of \qty{1.4}{\per\day}.
(This only represents `class 3' events, \ie{} those with the highest confidence of being real impacts.)
Since the \gorid{} sensor has a gathering power of about 15 times that of \heos{},
it effectively detected a 30 times lower flux of clustered impacts than \heos{}.\footnote{
  This discrepancy increases by a factor of \num{\sim2} if wall impacts are considered for \gorid{},
  due to a doubling of its geometrical factor.
  The discrepancy decreases by 33\% if wall impacts are considered for both instruments.
}
If the clusters detected by \heos{} and \gorid{} stem for the most part 
from the same phenomenon (presumably, the magnetospheric swarms), 
the discrepancy may be explained by the following reasons:
\begin{itemize}
\item The \heos{} sensor was more exposed to the anisotropic flow of swarms than \gorid{},
which both instruments indicated to be coming preferentially from the Earth apex:
\heos{} pointed straight towards the Earth apex for 40\% of the observation time.
\gorid{}, on the other hand, was only marginally exposed to the apex direction, 
as discussed in Section~\ref{SECT:Anisotropy}.
\item The \heos{} sensor was more sensitive than \gorid{}, and thus could detect more particles.
The particle mass detection threshold of the \heos{} detector is indeed reported to be 
one order of magnitude lower than that of \gorid{} at the same velocity:
$m_{\mathrm{min,HEOS}}\!=\!\qty{1.2e-15}{\gram}$ and 
$m_{\mathrm{min,GORID}}\!=\!\qty{1.5e-14}{\gram}$,
both at $v_{\mathrm{imp}}\!=\!\qty{10}{\km\per\s}$ \citep{Dietzel1973heos,Goller1989calibration}.
Particle masses derived from \heos{} measurements indicate that swarm particles are
partially below the \gorid{} threshold (\FetalAbbr{}).
\item More swarms occurred at the location of the \heos{} satellite, 
which flew through the pole regions of the magnetosphere,
as opposed to \gorid{}, which orbited in the equatorial plane.
Or, more swarms occurred during the time of the \heos{} mission,
potentially as a result of varying geomagnetic activity (see Section~\ref{SECT:geom_activity}).
\end{itemize}

Given the above considerations, the rates as of \gorid{} and \heos{} are arguably reconcilable,
which upholds the notion that the clusters detected by both instruments could stem from the same phenomenon.

\subsubsection{Swarm progenitor candidates}
We have already mentioned the apparent deficiency of low-velocity, low-bulk-density meteors,
whose progenitors could be the source of the swarms, yet which might not have
been observed due to selection effects, as noted by \FetalAbbr{}.
Especially considering the large masses derived by \FetalAbbr{} for these progenitors of
\qty{10}{\gram} to \qty{1000}{\kilo\gram} (geom.\@ mean of \qty{5.2}{\kilo\gram}),
the lack of their detection as meteors poses a conundrum.
In addition, \citet{Mendis1981role,Mendis1984entry} showed 
that---while conceivable for smaller grains---such large meteoroids would have to
exhibit improbably small tensile strengths in order to electrostatically disrupt.
Rather than complete disruption, they suggest that irregularly shaped meteoroids would be 
electrostatically `eroded', as edges and protrusions are more readily 
chipped by the electrostatic tension \citep{Hill1981electrostatic},
thereby shedding the material to generate the swarms.
As only a fraction of the bulk mass could be shed by this erosion, however,
even larger meteoroids would be required to yield the swarm masses estimated by \FetalAbbr{}.

Similarly speculative, one might consider the possibility that it is not the 
disruption of a single massive progenitor that produces the swarms,
but rather the quasi-simultaneous electrostatic breakup of numerous micrometeoroids
within a certain volume of space.
Given the asserted correlation of swarm detection with geomagnetic activity 
(see Section~\ref{SECT:geom_activity}), it is conceivable that 
local intensity spikes of the plasma environment could cause the disruption
of micron- and millimetre-sized particles within a whole region of the magnetosphere
\citep[for a review of turbulences of the magnetosphere, see, \eg{}][]{Zimbardo2010magnetic}.
Although only two of the 15 swarms detected by \heos{} exhibited a clear substructure 
(hinting at multiple simultaneous breakup events, \FetalAbbr{}),
such a scenario would lift the high mass requirement for the swarm progenitors,
so that other candidates could be considered.

In that regard, Jupiter family comets (JFC) have been found via dynamical modelling
to be able to produce also micrometeoroids on prograde orbits with very low semi-major axis
and aphelia near \qty{1}{\astronomicalunit}, which would generate 
low-velocity near-apex meteor radiants \citep{Nesvorny2011dynamical}.
As cometary meteoroids, which are generally associated with low bulk densities,
these could constitute the missing meteoroid subpopulation,
that produce the magnetospheric swarms.
Moreover, their preferential approach direction from the Earth apex 
would be consistent with the swarms' apparent anisotropy.
Quantitatively, their occurrence is expected to be only secondary to those from the
likewise JFC-generated, higher-velocity helion/antihelion radiants (\qty{\sim30}{\kms}), 
as well as the yet-higher-velocity north/south apex radiants (\qty{\sim60}{\kms})
sustained by Halley-type and long-period comets \citep{Nesvorny2011dynamical},
which, however, might less readily produce swarms, 
due to their shorter residence times within the magnetosphere.

A population of grains that appear to exhibit similar dynamics
has been detected in head-echo radar meteor with the Arecibo Observatory (AO). 
These slow meteors (\qty{\sim15}{\km\per\s}) emerged when AO pointed near the Earth apex,
alongside higher rates of fast meteors from the north/south apex radiants 
\citep{Janches2003geocentric,Sulzer2004meteoroid,Janches2005observed}.
The range of particle sizes constituting the AO dataset is stated to be
\qtyrange[range-units=single,range-phrase=--]{0.5}{100}{\um},
yet the sensitivity of head-echo radar meteor observations to slow particles \qty{<30}{\kms} 
has been a matter of debate and is likely closer to the higher end of the stated range 
\citep{Fentzke2008semiempirical,Janches2014radar,Janches2015radar,Janches2017radar}.

The dynamics of the swarms may hint at a connection to another group of interplanetary dust grains,
namely, the \amet{}s \citep{Grun1980dynamics,Sommer2023alphameteoroids}.
These particles move on highly eccentric orbits with aphelia near \qty{1}{\au}, where
they thus exhibit low heliocentric velocities.
The Earth effectively overtakes these slow-moving particles,
such that they appear to be coming from around the Earth apex
with $v_{\infty}\,=\,$\qtyrange[range-units=single,range-phrase=--]{5}{20}{\km\per\s}.
The \amet{}s notably exhibit the dynamical properties demanded for swarm progenitors, that is, 
a preferred approach direction from the apex, as well as relative velocities below \qty{20}{\km\per\s}.
Yet, these micron-sized particles are only about 1--2 orders of magnitude more massive
than the swarm particles themselves.

\subsubsection{Other explanations} \label{SECT:noise}
At the time of the \heos{} mission, the phenomenon of SRM dust streams was not yet well-known,
which is why they were not considered as explanation for the clusters in the interpretation of \heos{} data.
As it was established that GEO-insertion SRM firings could create distinct dust formations in orbit around Earth,
\citet{Fechtig1984interplanetary} acknowledged that they could be another explanation for the observed swarms.
However, there are reasons why the \heos{}-detected clusters are unlikely to be SRM exhaust dust formations,
namely the altitude and the locations at which they occurred.
The swarms were observed at all altitudes up to \qty{60000}{\km}.
In SRM GEO insertions, the exhaust dust is expelled rather retrograde such that the 
maximum obtainable apogee of ejected particles is near the GEO altitude (about \qty{36000}{\km})
\citep{Mueller1985effects,Bunte2003detectability}.
On the other hand, the orbit of \heos{} was oriented such that the spacecraft reached its far-out apogee near the 
ecliptic north direction, causing it to spend only very limited time near the equatorial region.
(see Figure~\ref{fig:HEOS_orbit}).
The north polar region was in use at the time by Russian Molniya satellites.
However, those were launched by the Molniya-M launcher, which used a liquid upper stage
(notwithstanding the fact that SRM dust ejected in Molniya orbit insertions would
fail to reach a stable orbit).

It is conceivable that orbital perturbations could have caused SRM dust formations to drift from their origin region
to where the swarms were observed.
However, it is questionable whether these displacements could take place in the lifetimes of SRM dust streams,
which are limited by atmospheric drag (at most 1--2 months for particles \qty{<1}{\um} in size
\citep{Friesen1992results,Bunte2003detectability}).
Moreover, electrodynamic perturbations on Earth-orbiting dust particles tend to dissipate
their orbital energy, adding to the orbital decay due to atmospheric drag
\citep{Juhasz1997dynamics}.
Even if perturbations can meaningfully move GEO-insertion SRM dust to high altitude polar regions,
their stochastic nature would arguably lead to the dispersion of the dust clouds.

Lastly, the Earth-apex anisotropy of the \heos{}-detected swarms has no obvious explanation in the context of SRM dust,
similarly to the preference of local midnight hours for cluster detections by \gorid{}.

An entirely different explanation could be that the consistently detected clusters are not 
true particle impacts but rather noise events caused by an interaction 
of the impact ionization detectors with the Earth's magnetosphere.
Spurious detections caused by various environmental factors have been a notorious issue
for dust-counter-type instruments, 
requiring sophisticated data reduction techniques \citep[\eg][]{Poppe2011measurements},
or invalidating datasets outright \citep{Nilsson1966doubts}.
In order to distinguish true impacts from noise events, a coincidence criterion was used
with \heos{} and \gorid{}, requiring valid successive charge signals 
from different measurement channels.
With \heos{}, a two-way coincidence criterion (target and ion collector) was used,
which for instance, classified all events registered during a historically powerful 
solar storm of August~1972 as noise \citep{Dietzel1973heos}.
\gorid{} on the other hand, employed a three-way coincidence criterion 
(target charge, ion grid, and channeltron) to identify true impacts with considerable confidence 
\citep{Baguhl1993identification,Grun1995reduction}.
(Note that \gorid{} is identical to the Ulysses and Galileo detectors.)
A significant amount of events classified by \gorid{} as probably noise-induced appears to be linked
to an electrostatic interaction with the spacecraft as well as to the operation of its
plasma thrusters, as described in \citep{Drolshagen2001measurementsb}.
Note that usually only the highest-confidence true-impact events have been used in the referred-to 
analyses of the \gorid{} data.
Nevertheless, it is at least conceivable that a so-far unknown interaction of the impact-ionization-type detectors
with the magnetosphere's turbulent plasma environment may cause the clustered occurrence
of genuine-looking noise events, despite the scrutiny put in place.
Given that electromagnetic effects are evidently conducive to the occurrence of noise events,
the apparent correlation of cluster detections with geomagnetic activity (Section~\ref{SECT:geom_activity})
is arguably consistent with that proposition.

\subsection{Future investigation with \des} \label{SECT:future_study}
\begin{figure}[h!tb]
  \centering
  \includegraphics[width=.7\linewidth]{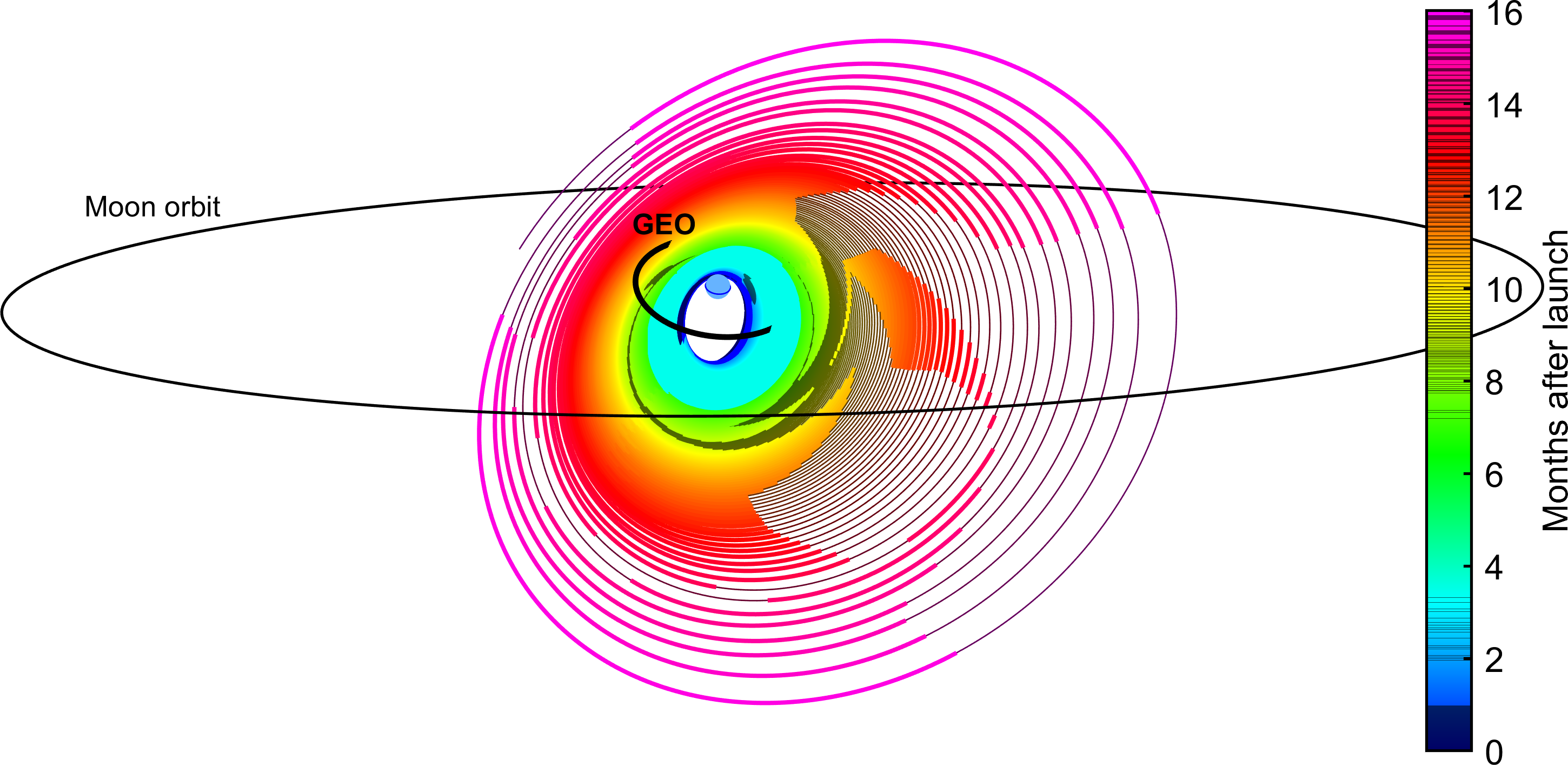}
  \caption{
    Trajectory of the initial phase of the \des{} mission,
    during which the spacecraft will gradually raise its orbit towards the Moon.
    Phases of thrusting and coasting are indicated by thick and thin lines, respectively.}
  \label{fig:destiny_trajectory}
\end{figure}

JAXA's \des{} mission is a planned small body science mission to active asteroid 3200 Phaethon,
scheduled for launch in 2025.
It carries the \des{} Dust Analyzer (DDA), a state-of-the-art dust detector that allows for the simultaneous
analysis of particles' dynamical (via charge-sensitive entry grids) 
and compositional (via an impact plasma mass spectrometer) information \citep{Simolka2023destiny}.
In addition to studying the dust environment of Phaethon,
DDA will conduct science operations during the entire nominal mission duration of 3 years,
to study interplanetary and interstellar dust \citep{Kruger2019modelling}.
For the first two years of the mission, the spacecraft revolves in Earth-bound orbits,
during which it gradually ascends from an initially GTO-like orbit all the way to the Moon's orbit
using its solar electric propulsion system, as shown in Figure~\ref{fig:destiny_trajectory}.
Eventually, \des{} will conduct a series of lunar gravity assists to escape
the Earth-Moon system and further advance towards Phaethon.

DDA will thus have extensive exposure to the near-Earth dust environment at all altitudes beyond LEO.
\des{} will pass through the magnetosphere at altitudes \qty{<10}{\earthRad} 
for roughly the first year of the mission,
and then traverse the plasma sheet region for several more months after that.
Unlike \heos{}, \des{} will remain within equatorial rather than the polar
regions of the magnetosphere, at an orbital inclination of \qty{\sim31}{\degree}.
The near-Earth mission phase of the mission will coincide with the solar cycle 25 maximum, 
which is expected for mid 2025.

The \des{} mission profile combined with the capabilities of the DDA instrument
represent a unique opportunity to further investigate the clustering phenomena 
previously reported by \heos{} and \gorid{}.
By recording a valid mass spectrum of a particle's impact plasma, 
DDA can unequivocally identify true particle impacts.
It could thus rule out the possibility of the clusters being spurious noise events.
The compositional analysis of swarm particles---should they be found---would then
give definitive proof of their natural or artificial origin.
Moreover, DDA's charge-sensing entrance grids may register larger swarm particles 
(reported by \heos{} to have masses up to \qty{e-11}{\gram} (\FetalAbbr{}),
corresponding to sizes of roughly \qty{\sim1}{\um}),
allowing for a direct measurement of their charge and velocity.
Such measurements would undoubtably be crucial for
understanding the swarms' formation mechanism.

Similarly to how we related the incidence rates of \heos{} and \gorid{} to each other,
we may tentatively estimate the incidence rate of swarm particles onto DDA
by converting the \heos{} and \gorid{} rates (see Section~\ref{SECT:Impact_rates})
to DDA rates using the relation of their respective geometrical factors
(see Table~\ref{tab:SensGeom_compareHeosGorid}).
This yields 133 swarm particle impacts per month onto DDA according to \heos{} rates
and 4.2 impacts per months according to \gorid{} rates (for the time within the magnetosphere).
Identifying the reason(s) for this discrepancy (as speculated upon in 
Section~\ref{SECT:Impact_rates}) would be a key aspect of the investigation.
It's important to note that this simple calculation ignores the differences
in spacecraft orbits and instrument pointings.

\section{Conclusion}
We have provided an overview of the various reports of clustered impact events 
given by dust-counter-type instruments in near-Earth space.
The clustered impacts detected by the \heos{} and \gorid{} impact ionization detectors within 
the magnetosphere and at GEO, respectively,
represent a peculiar phenomenon, which have been interpreted as transient particle swarms 
created either by the breakup of meteoroids, or by GEO spaceflight activities.
We have assessed the possibility that the swarms observed by both instruments 
are caused by the same phenomenon, which is indicated by commonalities
regarding their encounter durations and their anisotropy.
Remarkably, we have found that in both cases their occurrence was correlated
with high geomagnetic activity,
which is consistent with the notion that the swarms
are caused by the electrostatic breakup of meteoroids proposed by \FetalAbbr{},
but could also be interpreted as in favour of the clusters being spurious,
plasma-environment-induced noise events.
While the \gorid{}-detected clusters might also stem from the electrostatic breakup 
of SRM slag particles in GTO-like orbits \citep[as speculated by][]{Graps2007geo},
this scenario can hardly explain the \heos{}-detected swarms,
which were observed far from the equatorial plane and at altitudes of up to \qty{60000}{\km}.

Lacking the proper dynamical and compositional characterization of dust particles---which 
both \heos{} and \gorid{} were unable to provide---the nature of the swarms remains elusive.
The upcoming \des{} mission, however, will carry a sophisticated dust analyser instrument (the DDA),
which will offer these capabilities.
Considering the mission's prolonged orbit raising phase through the magnetosphere, 
\des{}/DDA poses a unique opportunity to finally unveil the origin of the putative particle swarms.

In light of the new insights presented here, it may be worth to revisit in particular the 
\gorid{} data, and analyse the timings and locations of cluster detections more thoroughly
with respect to geomagnetic activity and region within the magnetosphere.
Exploring dust counter data retrieved from the magnetospheres of Jupiter and Saturn for 
similar swarms could provide additional avenue for investigating this phenomenon.
While most observed clusters there were linked to dynamic nano-dust streams from the 
moons Io and Enceladus, there have also been reports of impact clusters of unresolved origin
seen by the Cassini Cosmic Dust Analyzer (CDA) in orbit around Saturn \citep{Hsu2011cassini}.

Finally, new insights about electrostatic effects in cosmic dust,
in particular the `electrostatic lofting' of grains from dusty surfaces 
suspended in a plasma or UV light \citep[\eg][]{Wang2016dust}, 
could help shed light on the formation mechanism of the magnetospheric swarms.

\vspace{10mm}
\subsection*{Acknowledgement}
I thank Simon Green and Tony McDonnell for their assistance
in acquiring valuable literature about the \gorid{} results,
as well as three anonymous reviewers for their constructive comments.
Partial support by the German Aerospace Center (DLR, grant no. 50OO2101)
is gratefully acknowledged.